\documentclass[12pt,twoside]{article}
\usepackage{xrb2007}

\def\nh{N$_\mathrm{H}$}
\def\cm2{cm$^{-2}$}

\def\integral{{\it{INTEGRAL}}}
\usepackage{graphicx,epsfig,epsf}
\usepackage{natbib}

\begin{document}

\session{Obscured XRBs and INTEGRAL Sources}

\shortauthor{Rodriguez \& Bodaghee}
\shorttitle{First 4 years of \integral\ HMXBs}

\title{The Galactic population of HMXBs as seen with \integral\ during its four
first years of activity}
\author{J\'er\^ome Rodriguez}
\affil{Laboratoire AIM, 
DAPNIA/SAp, F-91191 Gif-sur-Yvette, France}
\author{Arash Bodaghee}
\affil{ISDC, 16 chemin d'Ecogia, CH-1290, Versoix Switzerland}

\begin{abstract}
We collected the parameters (position, absorption, spin, orbital 
period, etc..), when known, of all Galactic sources detected by \integral\ 
during its four first years of activity. We use these parameters 
to test theoretical predictions. For example, it is clear that HMXBs tend 
to be found mostly in the tangential direction of the Galactic arms, 
while LMXBs tend to be clustered in the Galactic bulge. We then focus on 
HMXBs and present two possible new tools, in addition to the well-known 
``Corbet-diagram'', to distinguish between Be-HMXBs and Sg-HMXBs.
\end{abstract}

\section{Introduction}
Over its four first years of activity, \integral\ has detected about
300 known sources and discovered about 200 in the 20--100~keV range 
\citep[e.g.][]{bird07}. The sources discovered 
by \integral\ will hereafter be refered to as ``IGRs''\footnote{ 
An updated list of IGRs can be found at  
http://isdc.unige.ch/$\sim$rodrigue/html/igrsources.html}.
Amongst these objects 16\% are High Mass X-ray Binaries (HMXB), 16\% are 
Low Mass X-ray Binaries (LMXB) and a large fraction (26\%) still lacks 
a proper identification \citep{bodaghee07}.  Note that 
recent Swift follow-ups of IGRs have allowed the identification of several 
sources, either as AGN \citep[e.g.][]{landi07,rodriguez08}, or, mainly 
for those lying close to the Galactic plane, as probable 
X-ray binaries \citep{rodriguez08}. \\
\indent Unveiling the nature of sources of unknown type has a great 
importance. This allows us to study more precisely the source populations of 
the Galaxy, and thus has great implications on the scenario of source 
evolution, and on the evolution of the Galaxy. In the same manner populations
of extra galactic sources will have an impact on the evolution 
of galaxies in general and therefore have implications on cosmology. 
More generally, studying and understanding the nature of high energy sources
will permit us to study the physics of emitting processes. \\
\indent Identifying the nature of a given source is not an easy task, though.
A secure identification is usually obtained through the discovery of the 
infrared or optical counterpart. This first requires that the X-ray position
is refined to a few  arcsec accuracy, with either {\it{XMM-Newton}} or 
{\it{Swift}}, and in crowded regions of the sky, sub-arcsec accuracy 
(with {\it{Chandra}} then) is needed. This approach has  proven to 
be successful in a certain number of cases \citep[e.g.][]{tomsick06,chaty08}.
It is, however, difficult to obtain observing time with X-ray satellites, and 
even fine position is sometimes not sufficient to determine the true 
counterpart. \\
\indent Other parameters should thus be examined and may be used to identify 
 a given source. For example the X-ray spectral and temporal behavior can help.
Based on the X-ray spectra of IGR J16320$-$4751, 
\citet{rodriguez03,rodriguez06} 
suggested the source was most likely a supergiant (Sg)
HMXB with a neutron star primary. This 
was later confirmed by the detection of coherent pulsations in the 
X-ray flux from the source \citep{lutovinov05}. The HMXB and wind accreting 
nature of the source have recently been confirmed by \citet{chaty08} through 
infrared spectroscopy. Similarly, \citet[and these proceedings]{prat07} studied
the X-ray behavior of IGR J19140+0951, also suspected to be a Sg-HMXB 
\citep{rodriguez05}. Folding the spectral parameters on the known orbital
period, \citet{prat07} could first confirm the HMXB nature of the source 
(it was independently discovered to be a B0.5I HMXB), and 
deduce some parameters of the system such 
as the inclination angle, and the mass loss rate of the companion. These 
examples show that X-ray analysis can be a secure mean to identify the nature 
of sources. \\
\indent In an in-depth study \citet{bodaghee07} 
have collected all known parameters of all sources detected by \integral\ to 
analyze whether correlations exist between those parameters and the type of
source. This approach could permit them to test theoretical models, and search
for possible new tools to determine the source type. 
In the following, we present the results obtained for the Galactic sources. 
We first show that the position of a source on the plane of the sky, and the 
orbital period  can  give a first ``flavor'' of its nature. 
We then focus more particularly on HMXBs, and point out the importance 
of the absorption \nh\ in understanding those objects.
We show that in addition to the spin period vs. orbital period 
diagram (the so-called ``Corbet-diagram'' \citep{corbet84}), 
 two new tools can be used to discriminate amongst the nature of HMXBs.  

\section{LMXBs vs. HXMBs: positions and orbital periods}
\begin{figure}[htbp]
\epsfig{file=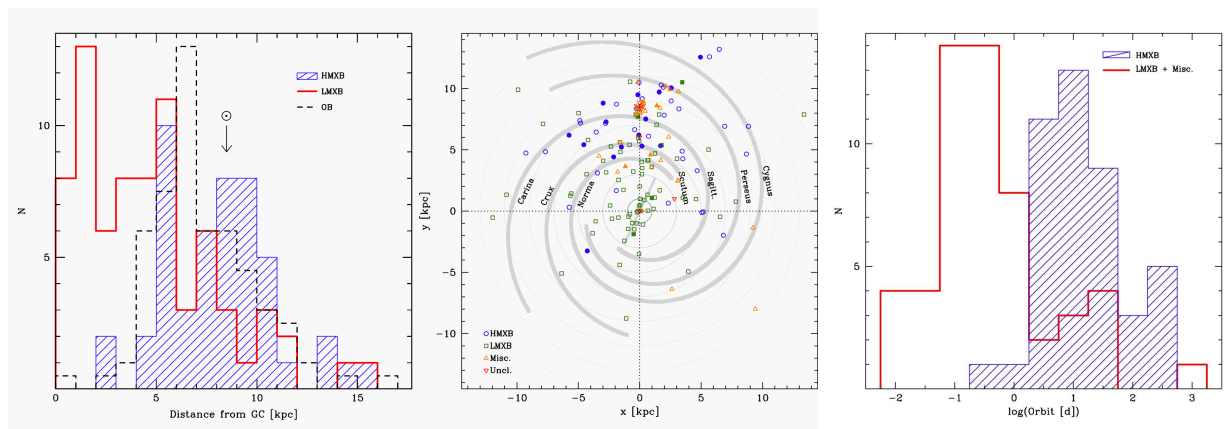,width=\columnwidth}
\caption{{\bf{Left :}} Distribution of Galactocentric distances of HMXBs and 
LMXBs. {\bf{Middle :}} Galactic distribution of HMXBs, LMXBs and other sources
superimposed on a 4-arm Galactic spiral model. {\bf{Right :}} Distribution 
of orbital periods of LMXBs and miscellaneous sources (e.g. CVs) and HMXBs.}
\label{fig:pos}
\end{figure}
It is evident from Fig. \ref{fig:pos} that a dichotomy between LMXBs and 
HMXBs exist. Probably the most obvious one concern the distribution of 
orbital periods. LMXBs have period mostly shorter than 1~d, while HMXBs
have much longer periods. This clearly is a results of the different histories
of the two populations of sources: HMXBs are young systems who did not have 
much time to evolve, while LMXB are old and may represent the ultimate stage
of the evolution of binary systems. \\
\indent In terms of position, a clear difference can also be seen. HMXBs tend 
to be found close to the Galactic arms \citep[see also e.g.][]{lutovinov05}, 
while LMXBs are more clustered in the Galactic bulge or off the plane. This 
galactic repartition of source again reflects the different histories of the 
two populations. LMXBs are old and have, thus, had time to migrate from their
formation site \citep[see, for example, the journey of XTE J1118+480 
across the Galaxy suggested by][]{mirabel}. They are therefore found where 
old globular cluster resides (in the bulge) or they had time to migrate 
off the plane (e.g. XTE J1118+480). HMXBs are, on the contrary, young, and 
did not have sufficient time to travel far from their formation sites. They 
are, hence, found close to star forming regions. Note that \citet{lutovinov05} 
found the distribution of HMXBs was slightly offset wrt the direction of 
the spiral arms. This could simply reflect the time delay between the evolution
of a system to become and HMXB and the change of the position of the 
Galactic arms induced by Galactic rotation.
\section{HMXBs: P$_{spin}$, P$_{orb}$ and \nh}
\begin{figure}[htbp]
\centering
\epsfig{file=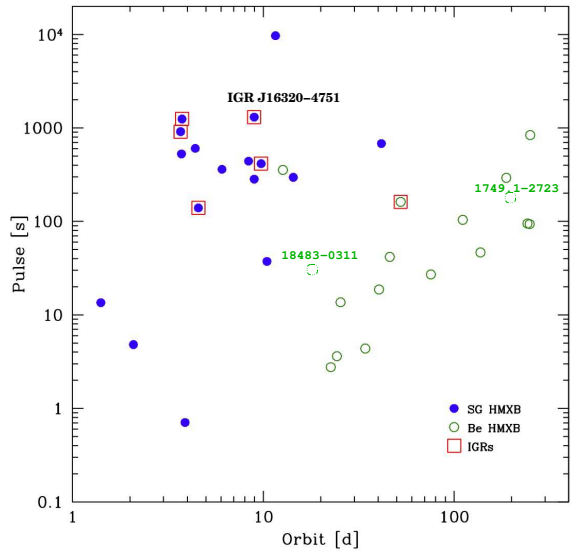,width=8cm}
\caption{Corbet-diagram of the HMXBs detected by \integral. The red squares
represent IGRs. Most IGRs are in the Sg-HMXB part of the plot. Note that 
in addition to the results from \citet{bodaghee07} two sources 
with recently found orbital and/or pulse periods have been added 
(IGR J18483$-$0311 and AX J1749.1$-$2723). 
IGR J16320$-$4751 early suspected to be a Sg-HMXB based on its X-ray 
behavior is also indicated.}
\label{fig:corbet}
\end{figure}

\begin{figure}[htbp]
\centering
\epsfig{file=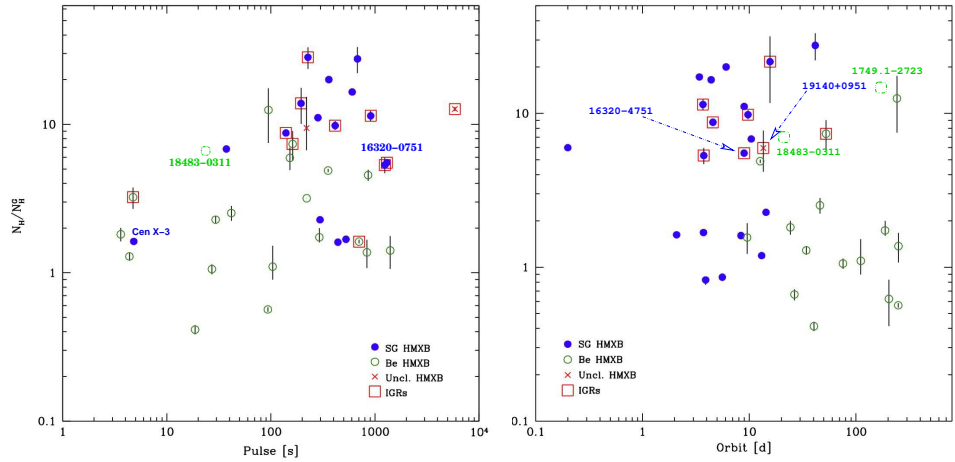,width=13cm}
\caption{``Bodaghee-diagrams'' of HMXBs. {\bf{Left :}} \nh\ vs. P$_{pulse}$.
{\bf{Right :}} \nh\ vs. P$_{orb}$. \nh\ is normalized by the value of the 
absorption on the line of sight as obtained from \citet{dickey}.}
\label{fig:bodaghee}
\end{figure}

While for extra-galactic IGRs are not more absorbed than previously 
known sources, this is not the case for Galactic sources. On average 
Galactic IGRs are four times more absorbed than pre-\integral\ sources.
Although not expected, this results is not astonishing. \integral/ISGRI is 
sensitive above 20~keV in a spectral domain immune to absorption, and 
\integral\ has performed a deep survey of the Galactic center and plane, where
Galactic sources are mostly found. Most of the identified Galactic IGRs are
Sg-HMXBs.  In addition to enabling a  
non-biased view of Galactic high energy source populations to be 
obtained, \integral\  has allowed new populations to be unveiled: 
the Supergiant Fast X-ray Transients and ``super-absorbed'' Sg-HXMBs 
(e.g. Chaty et al. , Rahoui et al. , Sidoli et al. , Negueruela et al. these
proceedings).\vspace*{0.1cm}
\\
{\bf{a). P$_{spin}$ vs. P$_{orb}$, the Corbet Diagram}}\\
\indent  Fig. \ref{fig:corbet} represents the Corbet-diagram  of the HMXBs with
known orbital and pulse periods. 
A well known dichotomy between Sg-HMXBs and Be-HMXBs is clear. 
As an example, IGR J16320$-$4751 is represented on this diagram. 
This source, early suspected to be a Sg-HMXB \citep{rodriguez03,rodriguez06},
and recently confirmed so \citep{chaty08}, lies well within the Sg-HMXB
part of the plot. Two sources (AX J1749.1$-$2723 and IGR J18483$-$0311) whose
orbital and/or pulse periods have recently been found are added to the plot. 
Given their position in the diagram they are proposed as Be-HMXBs 
\citep{sguera07,karasev07}. It is remarkable that although \integral\ 
has almost doubled the number of Sg-HMXBs, the known fact that Be-HMXBs 
have longer orbital periods remains valid.\vspace*{0.1cm}
\\ 
{\bf{b) \nh\ vs. P$_{pulse}$ and \nh\ vs. $P_{orb}$: the 
``Bodaghee-diagrams''}}\\
\indent Given the different accretion processes (pure wind vs. dense equatorial disk), 
different orbital periods, one may think that both classes of HMXBs
should have distinct behaviors when comparing their pulse and orbital 
periods with \nh. First Be systems have longer orbital periods than 
Sg-HMXBs. Then, in Sg-HMXBs the compact object is always embedded 
in the strong wind from the companion (hence \nh\ may expected to  
be higher), while in Be-HMXB this should not be the case. 
The different structures of the winds
(the density of the wind is $\rho(r)\propto r^{-2}$ in Sg stars, while 
the equatorial wind of Be star has $\rho(r)\propto r^{-3}$ with r the 
radius) may also affect differently the value of the spins of the pulsar in
both class of systems. Fig. \ref{fig:bodaghee} represents the \nh\ vs. 
P$_{pulse}$ and \nh\ vs. P$_{orb}$ diagrams.

Although the dichotomy is less clear than on the Corbet-diagram 
(Fig. \ref{fig:corbet}), a certain discrimination is visible on both 
plots. Both family of sources tend to cluster on different parts of
the diagrams. To test the validity of those 
diagrams to discriminate amongst the two different types, we can 
use  IGR J16320$-$4751, and IGR J19140+0951, both confirmed Sg-HMXBs.
The first has \nh=$2\times10^{23}$~\cm2, P$_{pulse}$=1300~s, 
and P$_{orb}=8.96$~d. It clearly lies within a region of Sg systems. The latter
has \nh=$10^{23}$~\cm2 and  P$_{orb}=13.55$~d. It also lies close to a group 
of Sg systems. We also added AX J1749.1$-$2723 and IGR J18483$-$0311 
to the plots. The former clearly lies in a region of the plot populated by 
Be systems. This and the its position in the Corbet diagram render its 
identification as a Be system very likely. The latter is more a borderline
source, although it seems in all three diagrams to lie closer to the Be 
systems. This suggests it is a Be-HMXB. Note that the Sg-HMXB clearly lying 
in the Be-HMXB zone in Fig. \ref{fig:bodaghee} (left) is Cen X-3 
(P$_{pulse}$=4.81~s, normalized \nh=1.7) a Roche lobe overflow HMXB, 
hence similar to Be-HMXBs.

\section{Conclusions}
We inspected all known parameters of about 500 sources detected by \integral\
\citep{bodaghee07}, and focused here mainly on X-ray binaries, and HMXBs.
The main results of our study can be summarized as follows:\\
$\bullet$ HMXBs are clustered towards spiral arm tangents although 
an offset is visible.\\
$\bullet$ The distribution of HMXBs follows star forming regions.\\
$\bullet$ LMXBs are mostly found in Galactic bulge,or at high 
latitude. \\
$\bullet$ A dichotomy between the orbital period of LMXBs/CVs and HMXBs 
exists and is confirmed even after four years of activity of \integral. \\
$\bullet$ \integral\ has discovered more than 200 IGRs\\
$\bullet$ The majority is unclassified but distribution follows 
Galactic population (although some recently identified are AGNs)\\
$\bullet$ 31\% of identified IGRs are HMXB, with large proportion of SG HMXB\\
$\bullet$ We showed that Sg-HMXBs and Be-HMXBs tend to occupy different 
regions of \nh\ vs. P$_{spin}$ and \nh\ vs. P$_{orb}$ diagrams\\
$\bullet$ This dichotomy may directly reflect differences related to 
the physics of accretion in those systems, and/or differences of the winds
of the companions.\\

\end{document}